# Inhomogeneous maps and mathematical theory of selection

## Georgy P. Karev

Management Systems Designers, Inc. ,

National Institutes of Health, Bldg. 38A, Rm. 5N511N, 8600 Rockville Pike, Bethesda, MD 20894, USA

Phone: (301) 451-6722; Fax: (301) 435-7794

E-mail:  karev@ncbi.nlm.nih.gov

**Abstract**.

In this paper we develop a theory of general selection systems with discrete time and explore the evolution of selection systems, in particular, inhomogeneous populations. We show that the knowledge of the initial distribution of the selection system allows us to determine explicitly the system distribution at the entire time interval. All statistical characteristics of interest, such as mean values of the fitness or any trait can be predicted effectively for indefinite time and these predictions dramatically depend on the initial distribution. The Fisher Fundamental theorem of natural selection (FTNS) and more general the Price equations are the famous results of the mathematical selection theory. We show that the problem of dynamic insufficiency for the Price equations and for the FTNS can be resolved within the framework of selection systems. Effective formulas for solutions of the Price equations and for the FTNS are derived. Applications of the developed theory to some other problems of mathematical biology (dynamics of inhomogeneous logistic and Ricker model, selection in rotifer populations) are also given. Complex behavior of the total population size, the mean fitness (in contrast to the plain FTNS) and other traits is possible for inhomogeneous populations with density-dependent fitness. The temporary dynamics of these quantities can be investigated with the help of suggested methods.

## 1. Introduction and background

### 1.1. The Fisher Fundamental theorem

According to the Darwinian theory, natural selection uses the genetic variation in a population to produce individuals that are adapted to their environment. A measure of an individual's ability to survive and reproduce is called fitness. The Fisher Fundamental theorem of natural selection (*FTNS*) states: "the rate of increase in fitness of any organism at any time is equal to its genetic variance in fitness at that time, except as affected by mutation, migration, change of environment and the effect of random sampling" ([8]).

A standard interpretation of the FTNS is that the rate at which natural selection acts upon a character distribution within a population is controlled by the variance of that character distribution ([30]). Many versions and special cases of the FTNS were proved within the framework of different exact mathematical models. One of the simplest models is as follows.

Let $N$ be the population size, $n_i$ be the number of alleles $A_i$, $p_i=n_i/N$ be the frequency and $w_i$ be the "fitness" of the $i$-th allele. Then

$$n_i` = w_i\, n_i. \qquad (1.1)$$

where primes denote one time step into the future and $w_i$ is the fitness of alleles $A_i$. The mean fitness is $E[w]=\sum p_i w_i$ and its variance is $Var[w]= E[w^2] - E^2[w]$. It follows from (1.1) that $p_i` = w_i p_i/E[w]$ and

$$\Delta E[w] \equiv \sum p`_i w_i - \sum p_i w_i = Var[w]/E[w]. \qquad (1.2)$$

So, within the framework of model (1.1) the FTNS (1.2) is a very simple mathematical assertion, which is not specific for genetic selection. Note that model (1.1) actually describes the dynamics of a subdivided (inhomogeneous) population with the fitness distributed over subpopulations; the subpopulations are composed of individuals that have the same fitness. Then assertion (1.2) is exactly the Li's theorem [21]: in a subdivided population the rate of change in the overall growth rate is proportional to the variance in the growth rates of the subpopulations.

Nevertheless, the actual biological content of the theorem, even if it were true mathematically, had been a subject of discussion in the literature for decades ([11], [6], [7], [3], etc.). Although Fisher himself noted that the FTNS only holds subject to important assumptions, he claimed that FTNS is an exact theorem that takes the supreme position among the biological sciences and indicates "the arrow of time", and compared it with the second law of thermodynamics.

In contrast with this claim, there are serious problems with interpretations of the standard FTNS. First of all, if the mean fitness is positive at any instant, then it increases according to (1.2) and hence the population size increases hyper-exponentially until the population becomes either homogeneous or infinite at a finite time moment (so called "population explosion", see, e.g., [16]). In reality and in some more realistic models the average fitness does not always increase. For example, the average fitness can decrease when selection acts on two linked loci that have epistatic effects on fitness (i.e., if the average effect of a substitution at one locus depends on the genotype frequencies at a second locus [3]). Below we show that the average fitness can behave in a very complex way if the fitness depends on the total population size, on the contrary with the standard FTNS.

Let us emphasize that the FTNS in the form (1.2) is a mathematical assertion that is valid within the framework of appropriate mathematical models of population dynamics but may be not applicable to some real situations. Sharp contradictions between reality and inevitable corollaries of the standard form of FTNS appear because its conditions fail. G. Price ([23], [25]) gave an explanation of the contradiction between Fisher's claim for generality and the limited scope of the usual interpretation of the FTNS (see also [11]). The total change $\Delta \bar{w}$ in average fitness $\bar{w}$ over time unit is $\Delta \bar{w} = \bar{w}` \mid En` - \bar{w} \mid En$ where $\bar{w} \mid En$ denote the average value of fitness under given environment $En$ and `denote one step in time. Hence,

$$\Delta \bar{w} = \{\bar{w}` \mid En - \bar{w} \mid En\} + \{\bar{w}` \mid En` - \bar{w}` \mid En\}.$$

Fisher called the first term in brackets the change in fitness caused by natural selection, and the second one as the change caused by deteriorations in the environment. The FTNS states that the change in fitness *caused by natural selection* is equal to the genic (see below s.3) variance in fitness. In such a formulation the FTNS is a true (at least for haploid populations in absence of epistasis) but not really a fundamental assertion [25]. As we have seen above, within the plain framework of model (1.1) it is a simple mathematical identity.

Let us emphasize that the only measure of the environment quality in the model is the fitness. Hence, deterioration of the environment results in the decrease of fitness and the changes in total fitness, which depends not only on natural selection but also on changes of the environment, which, in turn, can depend on the total population size. Fisher wrote: "An increase in number of any organism will impair its environment…The numbers must indeed be determined by the elastic quality of the resistance offered to increase in numbers, so that life is made somewhat harder to each individual when the population is larger, and easier when the population is smaller". These words actually describe and justify a well known transition from Malthusian models of free growing population to the population model with the size-dependent reproduction coefficient, such as logistic, Allee or other more complex models. It is a natural way of eliminating the "grave defect" of the FTNS.

Although the fitness is, perhaps, the most important characteristic of a population and individuals, the evolution of other particular traits is also of interest. The rate of change of a character, not the fitness itself, under natural selection depends on how closely the character is associated with fitness. Let $z$ be an arbitrary trait and $w$ be the fitness of individuals. The covariance equation

$$E[w] E[\Delta z] = Cov[w,z] \qquad (1.3)$$

was discovered independently by Robertson [29], Li [21], and Price [23]. It is easy to see that if $z=w$, then equation (1.3) turns into (1.2). It is the reason why the covariance equation is also known as "the second fundamental theorem of natural selection" [29].

The main difficulties with the correct formulation and application of FTNS appear when diploid populations are considered. Fitness of an individual depends on a huge number of characters and quantitative traits, which usually are influenced by a large number of loci. Theoretically, the population genetics of such traits could be described by gene frequencies. Practically, it is difficult to measure them; instead, the description can be done in terms of the distribution of characters. It is known (see, e.g., [30], [3]) that the expected genotypic value of an offspring does not depend in a simple way on those of its parents. With the aim to overcome this problem, Fisher suggested the least-square approximation of the genotypic value, $G_{ij} = \bar{G} + \gamma_i + \gamma_j + v_{ij}$, where $\gamma_i$ is called the average effect of $a_i$ and $v_{ij}$ is the dominance deviation, which should be chosen such that the dominance variation $\sigma^2_D = \sum_{i,j} v_{ij}^2 P_{ij}$ was minimized.

The part of the total genetic variance that can be accounted for the average effects of alleles is called the additive genetic variance, $\sigma^2_A = \sum_{i,j} (\gamma_i + \gamma_j)^2 P_{ij}$. If the population is in the Hardy-Weinberg proportions, i.e. $P_{ij} = p_i p_j$, then the following decomposition of the total genetic variance is valid: $\sigma^2_G = \sigma^2_A + \sigma^2_D$ (see [3], ch.1). It follows that the FTNS as it was initially formulated by Fisher is valid only in the limit case $\sigma^2_D = 0$, which is true, for example, for haploid populations or for additive fitness in the absence of dominance. The FTNS for diploid populations was discussed in detail in [7].

1.2. The Price equation

G. Price [24], [25] has shown in a general context of abstract selection theory that regardless of the particular model of population dynamics, the average fitness of a population increases at a rate proportional to the *total variance* in fitness. Both the FTNS and the covariance equation can be considered as special cases of the Price equation. Let the total population be subdivided into components or subpopulations of the sizes $n_i$, $i=1,…m$; let $p_i$ be frequency and $w_i$ be fitness of elements with index $i$, so that $n_i(t+1)=w_i(t)n_i(t)$. Now we assume that both the fitness and the trait can depend on time, so that $\Delta w$ and $\Delta z$ may not vanish. Then the 2$^{nd}$, or complete, Price equation is

$$\Delta E[z] = \{Cov[w,z] + E[w\Delta z]\} / E[w]. \qquad (1.4)$$

This equation shows the connection between the selection differential $\Delta E[z]$ for arbitrary trait and the fitness. In particular, if the character $z$ does not depend on $t$, i.e. $\Delta z=0$, then we get the 1$^{st}$ Price equation, which coincides with covariance equation (1.3). If we put $z=w$ in equation (1.4), then

$$\Delta E[w] = \{Var[w] + E[w\Delta w]\} / E[w].$$

If the fitness does not depend on time, i.e., $\Delta w=0$, then $\Delta E[w]=Var[w]/E[w]$, and it is the standard form of FTNS.

G. Price claimed [24] that his equation is the exact, complete description of evolutionary change under all conditions, in contrast to the FTNS and covariance equation, where the "environment" is fixed. It is worth noting that the Price approach actually did not take into account the effects of mutations; indeed, it follows from the equation $n_i(t)=w_i(t-1)n_i(t-1)=\prod_{s=0}^{t-1} w_i(s)n_i(0)$ that the subpopulation $n_i(t)$ is composed of individuals of type $i$ that are derived from type $i$ individuals at time 0. The more general "replicator-mutator" version of the Price equation was derived in [22].

The Price equation was applied not only to biological problems, such as evolutionary genetics, sex ratio, kin selection (see [27], ch.6, [5], [15], etc.), but also to social evolution [10], evolutionary economics [18], etc.

Lewontin [20] seems to be the first one to have noticed that the Price equation is not dynamically sufficient, i.e., it can not be used alone as a propagator of the dynamics of the model forward in time. Indeed, in order to calculate the dynamics of a mean trait with the help of the Price equation alone we need to solve the equation for the covariance, which in turn includes the moments of higher order. In general, this is impossible unless the higher moments are expressed in terms of lower moments (see, e.g., [1], [9]). The Price equation, being a mathematical *identity*, does not allow one to predict changes in the mean of a trait beyond the immediate response if only the value of covariance of the trait and fitness at this moment are known. For this, additional suppositions are required. To include the equation into the body of mathematical biology as a useful tool one should overcome the problem of dynamic insufficiency of this equation. Practically it means that some quantities in the Price equation should be calculated independently of others.

On the other hand, the following assertion is known about the asymptotic behavior of selection systems. If a limit distribution exists and is asymptotically stable, then it is concentrated in a finite number of points, which are the points of global maximum of an average reproduction coefficient on the support of the initial distribution. This "extremal principle" (see [31], [12], [13]) is a generalization of the Haldane principle [14]. So, based on the known results (the Price equation and the Haldane principle), the behavior of the selection systems can be predicted at the first time step and "at infinity".

Let us summarize the discussion.

A. The FTNS in its standard form, "the rate of increase in fitness of any organism at any time is equal to its *additive genetic, or genic variance* in fitness at that time", is only valid under substantial restrictions even within the framework of the simplest mathematical models (as Fisher wrote, "except as affected by mutation, migration, change of environment and the effect of random sampling");

B. The FTNS in the form "the rate of increase in fitness of any organism at any time is equal to its *total variance* in fitness at that time" is valid as a mathematical assertion for a broad class of models of inhomogeneous populations, where fitness does not change over time;

C. The FTNS in the form that allows dependence of the fitness on time is a particular case of the full Price equation, which is valid under quite general conditions;

D. The Price equations (and the FTNS) are not dynamically sufficient; they are mathematical identities within the framework of corresponding models and hence cannot be "solved" and cannot predict the population dynamics beyond the immediate response without additional suppositions.

In this paper we develop a theory of selection systems with discrete time and explore their evolution at the entire time interval where a global solution of the system is defined. We prove that the distribution of the system can be explicitly determined and computed at any time, therefore all statistical characteristics of interest, such as the mean values of the fitness or any trait can be computed effectively. In particular, the problem of dynamic insufficiency for the Price equations and for the FTNS can be resolved within the framework of selection systems if the initial distribution of the parameters is known. We derive explicit formulas for the solutions of the Fisher and Price equations under given initial distribution. We also derive the Haldane principle in explicit form for the considered class of selection systems.

The paper is organized as follows. The master model is formulated in s.2. The main features of our approach to inhomogeneous maps are demonstrated by the example of the simplest but important Malthusian model in s.3. Evolution of the main statistical characteristics of inhomogeneous populations described by the master model is explored in s.4. The results obtained in s.4 are specified in s.5 for the class of "self-regulated" maps that contains many well-known and wide-spreading models. The theory developed in ss. 4, 5 results the "solution" of the Price equations and the FTNS for selection systems (s.6). Other applications of general theory (inhomogeneous logistic and Ricker model, selection in rotifer populations) are considered in s.7.

## 2. Inhomogeneous maps as mathematical models for selection

Let us assume that a population consists of individuals, each of which is characterized by its own values of $n$ parameters $(a_1,\ldots a_n)=\mathbf{a}$. Here, we do not specify the vector-parameter $\mathbf{a}$, whose components may be arbitrary traits, e.g., $a_i$ could be the number of alleles of $i$-th gene, as in simple genetic models; we may think also of $\mathbf{a}$ as the entire genome. For the general case, we will denote $\{\mathbf{a}\}=A$.

Let $l(t,\mathbf{a})$ be the population density at moment $t$. Informally, $l(t,\mathbf{a})$ is the number of all individuals with a given vector-parameter $\mathbf{a}$; the subpopulation of all these individuals compose $\mathbf{a}$-clone. In general, the fitness of an individual depends on the individual vector-parameter $\mathbf{a}=(a_1,a_2,\ldots a_n)$ and on the "environment" that depends on time. Then in the next time instant

$$l(t+1,\mathbf{a})= w_t(\mathbf{a})\, l(t,\mathbf{a}) \tag{2.1}$$

where the reproduction rate $w_t(\mathbf{a})$ (fitness, by definition) is a non-negative function. The initial density $l(0,\mathbf{a})$ is supposed to be given.

Let $N(t) = \int_A l(t,\mathbf{a})d\mathbf{a}$ be the total population size; define

$$P_t(\mathbf{a})= l(t,\mathbf{a})/N(t), \tag{2.2}$$

to be the current probability density function (pdf). If $l(0,\mathbf{a})$ is given, then the initial pdf $P_0(\mathbf{a})$ is also given. It is important to note that, if $P_t(\mathbf{a}^*)=0$ for a particular $\mathbf{a}^*$ at some instant $t$, then $P_{t'}(\mathbf{a}^*)=0$ for all $t'>t$. Hence, selection system (2.1) describes the evolution of a distribution with a support that does not increase with time; it may be interpreted as the process of selection (see survey [13]).

We will show that, for a large class of models (2.1)-(2.2), the current pdf of the parameter, $P_t(\mathbf{a})$, can be computed if we suppose that the initial pdf of the parameter, $P_0(\mathbf{a})$, is known. Then, any term in the Price equation can be computed independently of others, and the problem of dynamical insufficiency disappears.

This class of models is defined by a certain condition on the reproduction rate $w_t(\mathbf{a})$. For the map (2.1) $w_t(\mathbf{a})>0$ and, hence, $w_t(\mathbf{a})=\exp(F_t(\mathbf{a}))$, where $F_t(\mathbf{a})=\ln[w_t(\mathbf{a})]$ is the "logarithmic reproduction rate". The exponential form of the reproduction coefficient is more appropriate in many cases for systems with discrete time [28]. Indeed, let $l(t+\Delta t,\mathbf{a})=l(t,\mathbf{a})(1+F_t(\mathbf{a})\Delta t)$ for a small time interval $\Delta t$. Then $dl(t,\mathbf{a})/dt=F_t(\mathbf{a})l(t,\mathbf{a})$. The difference analog of this equation is not $l(t+1,\mathbf{a})=F_t(\mathbf{a})l(t,\mathbf{a})$ but rather $l(t+1,\mathbf{a})=\exp[F_t(\mathbf{a})]l(t,\mathbf{a})$, because $l(t+\Delta t,\mathbf{a})=l(t,\mathbf{a})(1+F_t(\mathbf{a})\Delta t) \cong l(t,\mathbf{a})\exp[F_t(\mathbf{a})\Delta t]$; this equation coincides with the previous one if $\Delta t$ is taken as the time unit. In general, there is no one-to-one correspondence between difference and differential equations that describe the same system.

Taking into account that any smooth function of two variables $t$, $\mathbf{a}$ can be approximated by a finite sum of the form $\sum_i \varphi_i(\mathbf{a})g_i(t)$, where $\varphi_i$ depend on $\mathbf{a}$ only, and $g_i$ depend on $t$ only, we will suppose further that the fitness is of the form

$$w_t(\mathbf{a})=\exp[\sum_{i=1}^{n} \varphi_i(\mathbf{a})g_i(t)]. \tag{2.3}$$

Formula (2.3) defines the map from the set of all possible genotypes $\{\mathbf{a}\}=A$ to the set of corresponding fitness. Generally speaking, determination of this map is one of the central problems in biology. Within the framework of the master model (2.1)-(2.3), we take an individual fitness to depend on a given finite set of traits labeled by $i=1,…n$. The function $\varphi_i(\mathbf{a})$ describes the quantitative contribution of a particular $i$-th trait (or gene) to the total fitness, and $g_i(t)$ describe a possible variation of this contribution with time depending on the environment, population size, etc. Let us emphasize that we do not suppose that contributions of different traits are independent of one another, on the contrary, the evolution of this dependence is one of the central problems explored in this paper.

## 3. Malthusian inhomogeneous population model

To make clearer the features of our approach to investigation of inhomogeneous maps, let us consider the simplest but important example of the Malthusian version of inhomogeneous model (2.1). The model of population growth in absence of a density (or size) population regulation in stable environment is of the form

$$l(t+1,\mathbf{a}) = w(\mathbf{a})\, l(t,\mathbf{a}). \tag{3.1}$$

Let us collect together the main assertions about inhomogeneous Malthusian model (which follow from Theorem 1 below as a very particular case; see also [16], theorems 1.1, 2.5)).

For any measurable function $\varphi_t(\mathbf{a})$ defined on the probabilistic space $(A, P_t)$ (which can be considered as a random variable on this space) we will denote

$$E_t[\varphi_t] = \int_A \varphi_t(\mathbf{a}) P_t(\mathbf{a}) d\mathbf{a}.$$

*Let $P_0(\mathbf{a})$ be the initial pdf of the vector-parameter $\mathbf{a}$ for inhomogeneous map* (3.1). *Then*

1) *The population size $N_t$ satisfies the recurrence equation*

$$N_{t+1} = N_t\, E_t[w]\,; \tag{3.2}$$

2) *The current mean value of the fitness can be computed by formula*

$$E_t[w] = E_0[w^{t+1}]/E_0[w^t]; \tag{3.3}$$

3) *The current pdf $P_t(\mathbf{a})$ is given by the formula*

$$P_t(\mathbf{a}) = P_0(\mathbf{a})\, w^t(\mathbf{a})/E_0[w^t]. \tag{3.4}$$

The mean fitness $E_t[w]$ increases according to the FTNS (1.2) and hence the total population size may increase hyper-exponentially. The evolution of the mean fitness, population size and density dramatically depends on the initial distribution of the fitness even for this simplest model. Let us suppose for simplicity that the fitness itself is a distributed parameter and consider the evolution of current distribution of $w$ for a different initial pdf.

1) Let the initial pdf of $w$ be the $\Gamma$-distribution, $P_0(w=x)=s^k x^{k-1}\exp[-sx]/\Gamma(k)$ for $x\geq 0$, where $s, k>0$. Then $E_0[w^t]=\Gamma(k+t)/(\Gamma(k)s^t)$, and $P_t(w=x)$ is again the $\Gamma$-distribution with the parameters $s, k+t$; its mean is $E_t[w]=(k+t)/s$ and variance $Var_t[w]=(k+t)/s^2$.

Next, $N_{t+1}= E_t[w]N_t = (k+t)/s\, N_t$ and hence $N_t= N_0\Gamma(k+t)/(s^t\Gamma(k))$.

Indeed, if $P_0(w=x)$ is the $\Gamma$-distribution, then $E_0[w^t]=\Gamma(k+t)/(\Gamma(k)s^t)$. According to (3.4), $P_t(w=x)=s^k x^{k-1}\exp[-sx]/\Gamma(k)\{x^t/[\Gamma(k+t)/(\Gamma(k)s^t)]\}= s^{k+t}\, x^{k-1+t}\exp[-sx]/\Gamma(k+t)$ is again the $\Gamma$-distribution with the parameters $s, k+t$; its mean is $E_t[w]=(k+t)/s$, so $N_{t+1}=(k+t)/s\ N_t$ due to formula (3.2). (All other examples below can be proved in same way).

So, in this case the mean fitness increases linearly with time and the population size increases extremely fast, $N_t \sim t!/\ s^t$.

2) Let $P_0(w)$ be the log-normal distribution, $P_0(w=x)=\dfrac{1}{x\sigma\sqrt{2\pi}}\exp\{-\dfrac{(\ln x - m)^2}{2\sigma^2}\}$, $x>0$. Then $E_0[w^t]=\exp\{1/2 t^2\sigma^2+tm\}$, and $E_t[w]=E_0[w^{t+1}]/E_0[w^t]=\exp\{t\sigma^2 +1/2\sigma^2 +m\} \sim \exp\{\sigma^2 t\}$. The current distribution of the fitness is

$$P_t(w=x)=\dfrac{x^{t-1}}{\sigma\sqrt{2\pi}}\exp\{-\dfrac{(\ln x - m)^2}{2\sigma^2}-\dfrac{t^2\sigma^2}{2}-tm\}.$$

Next, $N_{t+1}=N_t E_t[f]=\exp\{t\sigma^2+1/2\sigma^2+m\}N_t$, so $N_t=N_0\exp\{(1/2\sigma^2)(t^2+2t)+mt\} \sim N_0\exp\{(1/2\sigma^2)t^2\}$.

We see that in this case the mean fitness increases exponentially with time, while the population grows as $\sim \exp\{(1/2\sigma^2)t^2\}$.

3) Let $P_0(f)$ be the Beta-distribution with parameters $(\alpha, \beta)$ in interval $[0,B]$; then $E_t[f]=B(\alpha+t)/(\alpha+t+\beta) \sim B$.

Next, $N_{t+1}= B(\alpha+t)/(\alpha+t+\beta)N_t$, so

$$N_t= N_0 B^t \dfrac{\Gamma(\alpha+t-1)\Gamma(\alpha+\beta)}{\Gamma(\alpha+\beta+t-1)\Gamma(\alpha)} \approx N_0 \dfrac{\Gamma(\alpha+\beta)}{\Gamma(\alpha)} B^t\ t^{-\beta}.$$

Hence, the fate of a population dramatically depends on the value of $B$: if $B\leq 1$, the population goes to extinction, if $B>1$, the size of the population increases indefinitely. In the case when $B=1$ the mean fitness tends to 1 and one could expect that the total population size would tend to a stable non-zero value in the course of time, but in fact the population becomes extinct at a power rate, $N_t \sim t^{-\beta}$.

4) Let $P_0(f)$ be the uniform distribution in the interval $[0,B]$. Then $E_0[f^t]=B^t/(t+1)$, hence $E_t[f]=B(t+1)/(t+2) \sim B$, and $N_t= N_0\ B^t /(t+1)$.

The fate of a population in this case also depends on the value of $B$: if $B>1$, the size of the population increases indefinitely, if $B\leq 1$, the population goes to extinct, $N_t \sim 1/t$.

In a particular case of Malthusian model (3.1) the fitness may be a function of a single selective trait, $w= w(a)$. The difference between mean values of the trait after and before selection (at $t$ time moment), $\Delta E[a]=E_{t+1}[a] - E_t[a]$, is known as the selection differential and is important characteristic of selection. The covariance equation and the Price equation show the relations between the selection differential and the fitness. As it has been shown above, the selection differential may evolve by different ways depending on the initial pdf. An opposite situation happens under the selection with strict truncation ([4], [33]). In this case, the fitness can be defined as follows: $w(a)=C$, if $a \leq B = const$, $w(a)=0$ if $a>B$.

Let $p_B = P_0(a \leq B)$, then $E_0(w^t) = C^t p_B$. Hence, $P_t(a) = P_0(a)/p_B$, for $a \leq B$, and $P_t(a)=0$ for $a>B$. This means that $P_t(a)$ is equal to the conditional probability $P_0(a)$ under condition that $a \leq B$, i.e. $P_t(a) = P_0(a)/P_0(a \leq B) \chi\{a \leq B\}$. Hence, the probability $P_t(a)$ does not change after the first selection step, $P_t(a) = P_1(a)$ for all $t>1$. Selection differential is $\Delta E_0[w] = E_0[w | a \leq B] - E_0[w]$ at the first selection step and $\Delta E_t[w] = 0$ for any $t>0$. Next, $E_t[w] = E_0[w^{t+1}]/E_0[w^t] = C$ and hence $N(t+1) = CN(t)$. Thus, the total size of the population increases (decreases) exponentially, $N(t) = N(0)C^t$, unless $C=1$. A more realistic model should take into account the population-size regulation.

Let $z(\mathbf{a})$ be any trait, i.e., a random variable on the space $(A, P_t)$. The key formula follows from (3.4):

$$E_t[z] = E_0[zw^t]/E_0[w^t]. \qquad (3.5)$$

This formula gives a way to circumvent the problem of dynamic insufficiency of the FTNS and the Price equations.

*Within the frameworks of inhomogeneous Malthusian map* (3.1), *under given initial pdf* $P_0(\mathbf{a})$ *the covariance equation,* $\Delta_t E_t[z] = Cov_t[z,w]/E_t[w]$, *has the solution* (3.5) *and all terms in the right hand side of the covariance equation can be computed explicitly*:

$Cov_t[z_t, w_t] = E_0[z w^{t+1}]/E_0[w^t] - E_0[z w^t]/E_0[w^t]$,

$E_t[w] = E_0[w^{t+1}]/E_0[w^t]$.

*The last equality gives the solution of the FTNS equation.*

The problem of dynamic insufficiency of the full Price equation is discussed and solved below for a more general model.

We have seen above that the mean fitness may increase indefinitely as a linear, exponential, power, etc. function of time depending on the initial distribution if its support is unbounded. Then the total population size also increases indefinitely and hyper-exponentially. If the fitness was initially distributed in a finite interval $[0,B]$ (as in examples 4 and 5) then its mean tends, in course of time, to the maximal possible value $B$. The fate of the entire population depends on the particular value of $B$: if $B>1$, the population increases asymptotically exponentially, if $B<1$, the population goes to extinction (if $B=1$, the population behavior may be different depending on the initial distribution). The exact values of the mean fitness and any other trait can be computed according to formulas (3.3), (3.5) at any instant if the initial distribution of the trait is known.

Remark that all results of this section can be extended to the models with "factorized" fitness of the form $w(\mathbf{a})=f(\mathbf{a})g(N_t)$. In particular, if the initial pdf of $f(\mathbf{a})$ is

1) *Gamma-distribution with the parameters (s, k), then $p_f(t;x)$ is again the Gamma-distribution with the parameters s, k+t;*

2) *Beta-distribution with parameters (α, β) then $p_f(t;x)$ is again the density of Beta-distribution with parameters (α+t, β).*

Evolution of other initial distributions was explored in [16], theorem 2.5. More general results are given in the next section.

## 4. Evolution of the main statistical characteristics of inhomogeneous maps

Let us return to master model (2.1), (2.3). The following main Theorem 1 shows that the model can be reduced to a non-autonomous map on $I \subseteq \mathbf{R}^1$ and completely explored. Let us denote

$$K_t(\mathbf{a}) = \prod_{k=0}^{t} w_k(\mathbf{a}) = \exp(\sum_{i=1}^{n} \varphi_i(\mathbf{a})G_i(t)) \qquad (4.1)$$

where $G_i(t) = \sum_{s=0}^{t} g_i(s)$. It is easy to see that $w_t K_{t-1} = K_t$ and $l(t+1,\mathbf{a}) = K_t(\mathbf{a})\, l(0,\mathbf{a})$.

We could think of the function $K_t(\mathbf{a})$ as the reproduction coefficient for the [0,t]-period or, for short, *t-fitness*. Let us note, that sometimes the functions $g_i(t)$ and hence $G_i(t)$ can be well defined not for all $0<t<\infty$, but only for $0<t<T$, where $T$ is a certain finite time moment. Accordingly, all assertions below can be valid only for $t<T$. Below we do not specify this condition if it is not necessary.

Let us denote $\vartheta=(\varphi_1, \varphi_2, \ldots \varphi_n)$ and let $\boldsymbol{p}(t;\vartheta)$ be the pdf of the random vector $\vartheta$ at $t$ moment, i.e. $\boldsymbol{p}(t;x_1, \ldots x_n) = P_t(\varphi_1=x_1,\ldots \varphi_1=x_1)$. The master model defines a complex transformation of the distribution $P_t(\mathbf{a})$ or, equivalently, the pdf $\boldsymbol{p}(t;\vartheta)$ in course of time.

Let $\boldsymbol{\lambda}=(\lambda_1,\ldots \lambda_n)$; denote

$$M_t(\boldsymbol{\lambda}) = \int_A \exp(\sum_{i=1}^{n} \lambda_i \varphi_i(\mathbf{a}))P(t;\mathbf{a})d\mathbf{a} = \int_{R_n} \exp(\sum_{i=1}^{n} \lambda_i x_i)\boldsymbol{p}(t;x_1,\ldots x_n))dx_1,\ldots dx_n \quad (4.2)$$

the moment generation function (mgf) of the pdf $\boldsymbol{p}(t;\vartheta)$ of the random vector $\vartheta$.

The initial pdf $p(0;\vartheta)$ is supposed to be given. The mgf of the initial distribution, $M_0(\lambda)$, is crucially important for the theory developed below. For example, $E_0[K_t]$ can be easily computed with the help of $M_0(\lambda)$:

$E_0[K_t] = M_0(\mathbf{G}(t))$

where we denoted $\mathbf{G}(t)=(G_1(t),\ldots G_n(t))$.

**Theorem 1.** *Let $P_0(\mathbf{a})$ be the initial pdf of the vector-parameter $\mathbf{a}$ for inhomogeneous map* (2.1), (2.3). *Then*

1) *The population size $N_t$ satisfies the recurrence equation*

$N_{t+1} = N_t E_t[w_t]$ ;

*and can be computed by the formula*

$N_t = N_0 E_0[K_{t-1}] = N_0 M_0(\mathbf{G}(t-1))$

2) *The current pdf $P_t(\mathbf{a})$ satisfies the recurrence equation*

$P_{t+1}(\mathbf{a}) = P_t(\mathbf{a}) w_t(\mathbf{a}) / E_t[w_t]$

*and can be computed by the formula*

$P_t(\mathbf{a}) = P_0(\mathbf{a}) K_{t-1}(\mathbf{a}) / E_0[K_{t-1}]$

3) *Let $\psi(\mathbf{a})$ be a random variable on the space $(A, P_t)$. Then*

$E_t[\psi] = E_0[\psi K_{t-1}] / E_0[K_{t-1}]$.

**Proof.** Rewriting equations (2.1), (2.3) as $l(t+1,a)/l(t,a) = \exp(\sum_{i=1}^{n} \varphi_i(a) g_i(t))$, we see that

$l(t,a) = l(0,a) \exp[\sum_{i=1}^{n} \varphi_i(a) \sum_{s=0}^{t-1} g_i(s)] = l(0,a) K_{t-1}(a)$

Then $N_t = \int_A l(t,a) da = N_0 E_0[K_{t-1}]$ and

$P_t(a) = l(t,a)/N_t = P_0(a) K_{t-1}(a) / E_0[K_{t-1}(a)]$.

Next, integrating over $a$ the equality $l(t+1,a) = w_t(a) P_t(a) N_t$ we obtain

$N_{t+1} = N_t E_t[w_t]$.

The mean value of a r.v. $\psi(\mathbf{a})$ at $t$ instant is

$E_t[\psi] = \int_A \psi(\mathbf{a}) P_t(\mathbf{a}) d\mathbf{a} = (\int_A \psi(\mathbf{a}) K_{t-1}(\mathbf{a}) P_0(\mathbf{a}) d\mathbf{a})/E_0[K_{t-1}] = E_0[\psi K_{t-1}]/E_0[K_{t-1}]$.

In particular,

$E_t[w] = E_0[w K_{t-1}]/E_0[K_{t-1}] = E_0[K_t]/E_0[K_{t-1}]$.

So, $P_{t+1}(\mathbf{a})/P_t(\mathbf{a}) = [K_t(\mathbf{a})/K_{t-1}(\mathbf{a})] / \{ E_0[K_t]/E_0[K_{t-1}]\} = w_t / E_t[w_t]$.

Q.E.D.

Denote $k(\mathbf{a}) = \overline{\lim}_{t \to \infty}[1/t \sum_{s=1}^{t} (\sum_{i=1}^{n} \varphi_i(\mathbf{a}) g_i(s))]$ the average reproduction rate of an **a**-clone. Then $K_t(\mathbf{a}_1)/ K_t(\mathbf{a}_2) = \exp[\sum_{i=1}^{n} \sum_{s=0}^{t} \varphi_i(\mathbf{a}) g_i(s)] \approx \exp[t(k(\mathbf{a}_1) - k(\mathbf{a}_2))]$.

The following corollary helps to understand the evolution of distribution and explains the Haldane principle within the frameworks of the master model.

Let $P_0(\mathbf{a}_2) > 0$. Then

$P_t(\mathbf{a}_1)/P_t(\mathbf{a}_2) = P_0(\mathbf{a}_1)/P_0(\mathbf{a}_2) [K_{t-1}(\mathbf{a}_1)/ K_{t-1}(\mathbf{a}_2)] \approx$
$P_0(\mathbf{a}_1)/P_0(\mathbf{a}_2) \exp[t(k(\mathbf{a}_1) - k(\mathbf{a}_2))]$. (4.3)

Hence, the evolution of a heterogeneous population leads to an (exponentially fast) replacement of individuals with smaller values of $k(\mathbf{a}_1)$ by those with greater values of $k(\mathbf{a}_2)$, even though the fraction of the latter in the initial distribution was arbitrarily small. Let **a*** be a point of global maximum of $k(\mathbf{a})$ and $P_0(\mathbf{a}^*) > 0$; then $k(\mathbf{a}) < k(\mathbf{a}^*)$ implies $P_t(\mathbf{a}) \to 0$. So, any stationary or limit distribution (if exists) should be concentrated in the set of points of global maximum of the average reproduction rate $k(\mathbf{a})$ on the support of the initial distribution. This version of the Haldane principle was established in [31].

The dynamics of the distribution on finite times is also of interest and, perhaps, is even more important for applications then detailed mathematical description of limit distributions (which may not exist even for plain population models). In the framework of the master model, the general case of the complete description of $P_t(\mathbf{a})$ is given by Theorem 1.

Dynamics and transformation with time of particular initial distributions is of interest for many practical problems. The following lemma is a key for investigation of the evolution of distributions.

**Lemma 1.** $M_t(\lambda) = M_0(\lambda + \mathbf{G}(t-1)) / M_0(\mathbf{G}(t-1))$.

**Proof.**

$M_t(\lambda) = \int_A \exp(\sum_{i=1}^{n} \lambda_i \varphi_i(\mathbf{a})) P_t(\mathbf{a}) d\mathbf{a} =$

$\int_A \exp(\sum_{i=1}^{n} \lambda_i \varphi_i(\mathbf{a})) K_{t-1}(\mathbf{a}) / E_0[K_{t-1}] P_0(\mathbf{a}) d\mathbf{a} =$

$\int_A \exp(\sum_{i=1}^{n} (\lambda_i + G_i(t-1)) \varphi_i(\mathbf{a})) / E_0[K_{t-1}] P_0(\mathbf{a}) d\mathbf{a} = M(0; \lambda + \mathbf{G}(t-1)) / M_0(\mathbf{G}(t-1))$.

Q.E.D.

Let us start from the simplest but important case when the r.v. $\varphi_i$ are independent at the initial instant.

**Theorem 2.** *Let the random variables $\{\varphi_i, i=1,\ldots n\}$ be independent at the initial instant and have the initial distributions $p_0^i(x_i)$ with the mgfs $M_0^i(\lambda_i)$, so that $\boldsymbol{p}_0(x_1, \ldots x_n) = \prod_{i=1}^{n} p_0^i(x_i)$ and $M_0(\lambda_1,\ldots \lambda_n) = \prod_{i=1}^{n} M_0^i(\lambda_i)$. Then for any $t \in [0,T)$ the r.v.-s $\varphi_i$ are independent, their distributions $p_t^i(x_i)$ have the mgf $M_t^i(\lambda_i) = M_0^i(\lambda_i + G_i(t))/M_0^i(G_i(t))$ and*
$$M_t(\lambda_1,\ldots \lambda_n) = \prod_{i=1}^{n} M_0^i(\lambda_i + G_i(t))/M_0^i(G_i(t)).$$

**Proof.**

According to Lemma 1, $M_t(\lambda) = M_0(\lambda + \Phi(t-1)) / M_0(\mathbf{G}(t-1)) =$
$$\int_A \exp\left[\sum_{i=1}^{n} (\lambda_i + \Phi_i(t))x_i\right] \prod_{i=1}^{n} p_i(0;x_i)\, dx_1,\ldots dx_n / M_0(\mathbf{G}(t-1)) =$$
$$\prod_{i=1}^{n} \{M_0^i(0;\lambda_i + \Phi_i(t))/M_0^i(0; \Phi_i(t))\}.$$

Q.E.D.

The evolution of the pdf of the random vector $\vartheta = (\varphi_1, \varphi_2, \ldots \varphi_n)$ in general case of correlated r.v.-s. $\{\varphi_i, i=1,\ldots n\}$ is of great practical interest because it helps to explore the dynamics of an inhomogeneous population depending on correlations between the random variables $\varphi_i(\boldsymbol{a})$.

Let us call a class $S$ of probability distributions of the random vector $\vartheta = (\varphi_1, \ldots \varphi_n)$ *invariant with respect to the model* (2.1), (2.3), *if $p(0,\vartheta) \in S \Rightarrow p(t,\vartheta) \in S$ for all $t$*.

Let $\boldsymbol{MS}$ be the class of moment-generating functions for distributions from the class $S$. The criterion of invariance immediately follows from Lemma 1.

*A class $S$ of pdf is invariant with respect to model* (2.1), (2.3) *if and only if*

$M_0(\lambda) \in \boldsymbol{MS} \Rightarrow M_0(\lambda + \mathbf{G}(t)) / M_0(\mathbf{G}(t)) \in \boldsymbol{MS}$ *for all $t$.*

We can prove with the help of this criterion that many important distributions are invariant with respect to model (2.1), (2.3). Let us recall some definitions (see [19]).

A random vector $X = (X_1,\ldots X_n)$ has a multivariate normal distribution with the mean $EX = \boldsymbol{m} = (m_1,\ldots m_n)$ and a covariance matrix $\boldsymbol{K} = \{c_{ij}\}$, $c_{ij} = \text{cov}(X_i, X_j)$ if its mgf is
$$M(\lambda) = E[\exp(\lambda^T X)] = \exp[\lambda^T \boldsymbol{m} + 1/2 \lambda^T C\lambda].$$

A random vector $X=(X_1,\ldots X_n)$ has a multivariate polynomial distribution with parameters $(k; p_1,\ldots p_n)$, if $P\{X_1=m_1,\ldots X_n=m_n\}=\dfrac{k!}{m_1!\ldots m_n!}p_1^{m_1}\ldots p_n^{m_n}$ for $\sum_{i=1}^{n}m_i=k$. The mgf of the polynomial distribution is $M(\lambda)=(\sum_{i=1}^{n}p_i\exp(\lambda_i))^k$.

A general class of *multivariate natural exponential distribution* is very important for applications; this class includes multivariate polynomial, normal, and Wishart distributions as particular cases.

A random *n*-dimension vector $X=(X_1,\ldots X_n)$ has multivariate natural exponential distribution (NED) with parameters $\theta=(\theta_1, \theta_2,\ldots \theta_n)$ with respect to the positive measure $\nu$ on $R^n$ if its joint density function is of the form

$$f_\theta(X)=h(X)\exp[X^T\theta-s(\theta)] \qquad (4.4)$$

where $s(\theta)$ is a function on parameters.

It is supposed that the *generating measure* $\mu(dX)=h(X)\nu(dX)$ is not concentrated on any affine hyperplane of $R^n$. The mgf of NED (4.4) is

$$M(\lambda)=E[\exp(\lambda^T X)]=\exp[s(\theta+\lambda)-s(\theta)]$$

and the cumulant generating function is $K(\lambda)=s(\theta+\lambda)-s(\theta)$.

For any NED, $\dfrac{\partial}{\partial\theta_i}s(\theta)=m_i$, and $\dfrac{\partial^2}{\partial\theta_i\partial\theta_j}s(\theta)=\mathrm{cov}(X_i,X_j)$.

**Theorem 3**. Let *us assume, that in the initial time moment the random vector* $\vartheta=(\varphi_1, \varphi_2,\ldots \varphi_n)$ *has*

i) *multivariate normal distribution with the mean vector* $m(0)$ *and covariance matrix* $K=(c_{ik})$. *Then at any moment t the vector* $\vartheta$ *also has the multivariate normal distribution with the same covariance matrix* $K$ *and the mean vector* $m(t)$, $m_i(t)=m_i(0)+1/2\sum_{k=1}^{n}[c_{ik}+c_{ki}]G_k(t-1)$;

ii) *multivariate polynomial distribution. Then at any moment t the vector-parameter* $a$ *has the multivariate polynomial distribution with parameters* $(k;p_1(t),\ldots p_n(t))$, *where* $p_i(t)=p_i\exp(G_i(t-1))/\sum_{j=1}^{n}p_j\exp(G_j(t-1))$.

iii) *multivariate natural exponential distribution on* $R^n$ *with moment generating function* (4.4) *and parameters* $\theta=(\theta_1, \theta_2,\ldots \theta_n)$. *Then at any moment t the vector* $\vartheta$ *also has* the multivariate NED with the parameters $\theta+G(t-1)$ and the moment generating function $\exp[s(\theta+\lambda+G(t-1))- s(\theta+G(t-1))]$.

**Proof.**

i) The mgf of the initial distribution of the vector $\vartheta$ is

$$M_0(\lambda_1,\ldots \lambda_n) = \exp(\sum_{i=1}^{n} \lambda_i\, m_i(0) + 1/2 \sum_{i,k=1}^{n} \lambda_i\, c_{ik}\, \lambda_k);$$

due to Lemma 1,

$$M_t(\lambda_1,\ldots \lambda_n) = M_0(\lambda+ \mathbf{G}(t-1)) / M_0(\mathbf{G}(t-1)) =$$

$$\exp[\sum_{i=1}^{n} (\lambda_i + G_i(t-1))m_i(0) + 1/2 \sum_{i,k=1}^{n} (\lambda_i + G_i(t-1))c_{ik}(\lambda_k + G_k(t-1)) -$$

$$- \sum_{i=1}^{n} G_i(t-1)m_i(0) - 1/2 \sum_{i,k=1}^{n} G_i(t-1)\, c_{ik}\, G_k(t-1)] =$$

$$\exp[\sum_{i=1}^{n} \lambda_i \{m_i(0)+1/2 \sum_{k=1}^{n} [c_{ik}+c_{ki}]\, G_k(t-1)\} + 1/2 \sum_{i,k=1}^{n} \lambda_i c_{ik} \lambda_k],$$

and this is the mgf of desired multivariate normal distribution.

ii) If the vector $\vartheta$ has the polynomial distribution at the initial time moment, then

$$M_0(\lambda_1,\ldots \lambda_n) = (\sum_{i=1}^{n} p_i\exp(\lambda_i))^k.$$

Due to Lemma 1

$$M_t(\lambda_1,\ldots \lambda_n) = (\sum_{i=1}^{n} p_i\exp(\lambda_i+ G_i(t-1)))^k / (\sum_{i=1}^{n} p_i\exp(G_i(t-1)))^k =$$

$$(\sum_{i=1}^{n} p_i(t)\exp(\lambda_i))^k$$

with $p_i(t)= p_i \exp(G_i(t-1))/ (\sum_{i=1}^{n} p_i\exp(G_i(t-1)))$, and this is the mgf of desired multivariate polynomial distribution.

iii) If the vector $\vartheta$ has at the initial time the multivariate natural exponential distribution with the mgf

$M_0(\lambda_1,\ldots \lambda_n) = \exp[s(\theta+\lambda)-s(\theta)]$, then

$M_t(\lambda_1,\ldots \lambda_n) = \exp[s(\theta+\lambda+G(t-1))-s(\theta)]/\exp[s(\theta+G(t-1))-s(\theta)] =$

$\exp[s(\theta+\lambda+G(t-1))- s(\theta+G(t-1))]$.

Q.E.D.

Theorem 3 states that multivariate normal, polynomial, Wishart', natural exponential distributions are invariant with respect to the master model. On the contrary, the multivariate uniform distribution is an example of distributions that are not invariant.

**Definition**. Let $S$ be a bounded Borel set in $R^n$ and $mesS$ be its Lebesgue measure. A random $n$-dimension vector $X=(X_1,...X_n)$ has a multivariate uniform distribution in $S$ if $mesS>0$ and its pdf is

$$f(X) = \begin{cases} 1/mesS, X \in S \\ 0, X \notin S. \end{cases}$$

The moment generating function of the uniform distribution in the case of rectangle $S=\{b_i \leq a_i \leq c_i, i=1,...n\}$ is

$$M(\lambda_1,...\lambda_n) = \{\prod_{i=1}^{n} (\exp(\lambda_i c_i) - \exp(\lambda_i b_i))\} / \{\prod_{i=1}^{n} (c_i - b_i) \prod_{i=1}^{n} \lambda_i\}.$$

Let the vector $\vartheta=(\varphi_1, \varphi_2, ... \varphi_n)$ has at the initial moment a multivariate uniform distribution in a rectangle. Then Lemma 1 implies that

$$M_t(\lambda_1,...\lambda_n) = \{\prod_{i=1}^{n} (\exp((\lambda_i + G_i(t-1))c_i) - \exp((\lambda_i + G_i(t-1))b_i))\} /$$

$$\{\prod_{i=1}^{n} (1+\lambda_i/G_i(t-1)) \prod_{i=1}^{n} (\exp(G_i(t-1)c_i) - \exp(G_i(t-1)b_i))\}.$$

So, at any moment $t>0$ the vector $\vartheta$ has not the uniform but the multivariate truncated exponential distribution in this rectangle.

### 5. Self-regulated inhomogeneous maps

The theory developed in s.4 for population models with the fitness of form (2.3) can be applied only if the time-dependent components $g_i(t)$ are known explicitly. As a rule, this is not the case for most interesting and realistic models where the time-dependent component should be computed according to the current population characteristics. For example, the logistic-type model accounting "deterioration of environment" when the population increases corresponds to the function $g(N_t)=1-N_t/B$ where $B$ is the carrying capacity; the *Ricker'* model corresponds to the function $g(N_t)= \lambda\exp(-\beta N_t)$.

In general, the population regulation ("change of environment due to the population' growth") may be defined not only by the total population size but also by the so-called "regulators", which are the averages over the population density:

$$S_i(t) = \int_A s_i(\mathbf{a})l(t,\mathbf{a})d\mathbf{a}, \qquad (5.1)$$

or

$$H_i(t) = \int_A h_i(\mathbf{a})P(t,\mathbf{a})d\mathbf{a} \qquad (5.2)$$

where $s_i(\mathbf{a})$, $h_i(\mathbf{a})$ are appropriate functions. For example, if $h(\mathbf{a})=s(\mathbf{a})$ is a biomass of an individual with parameter $\mathbf{a}$, then $H(t)$ is an average biomass, $S(t)$ is a total population biomass and the population growth rate may depend on $S(t)=N_t H(t)$. The total population size $N_t$ is also a regulator with $s(\mathbf{a})=1$.

Note that there exists a simple relation between the regulators (5.1) and (5.2):

$$\int_A s(\mathbf{a})l(t,\mathbf{a})d\mathbf{a} = N_t \int_A s(\mathbf{a})P(t,\mathbf{a})d\mathbf{a},$$

nevertheless, it may be useful for some situation to distinguish between the "density-dependent" (5.1) and "frequency-dependent" (5.2) regulators.

So, let us specify the theory developed in s.4 to the case of the model

$$l(t+1,\mathbf{a}) = l(t,\mathbf{a}) w_t(\mathbf{a}), \qquad (5.3)$$

$$w_t(\mathbf{a}) = \exp\left[\sum_{i=1}^{n} u_i(S_i(t))\varphi_i(\mathbf{a}) + \sum_{j=1}^{m} v_j(H_j(t))\psi_j(\mathbf{a})\right],$$

which we will refer to as the self-regulated inhomogeneous population model. Here the individual fitness $w_t(\mathbf{a})$ is *regulator-dependent*, i.e. may depend on some density-dependent and frequency-dependent regulators, $S_i(t)$ and $H_j(t)$. The initial distribution $l(0,\mathbf{a})$ of individuals over the vector-parameter $\mathbf{a}$ is supposed to be given. The main new problem is that the values of regulators are not given but should be computed at each point in time. Now $t$-fitness (4.1) is equal to

$$K_t(\mathbf{a}) = \prod_{k=0}^{t} w_k(\mathbf{a}) = \exp\left[\sum_{k=0}^{t} \left(\sum_{i=1}^{n} \varphi_i(\mathbf{a}) u_i(S_i(k)) + \sum_{j=1}^{m} \psi_j(\mathbf{a}) v_j(H_j(k))\right)\right].$$

Let $f(\mathbf{a})$ be a (measurable) function on $A$ and $\boldsymbol{\lambda}=(\lambda_1,\ldots \lambda_n)$, $\boldsymbol{\delta}=(\delta_1,\ldots \delta_m)$. For given initial distribution $P_0(\mathbf{a})$, introduce the functional $\mathbf{M}(f; \boldsymbol{\lambda}, \boldsymbol{\delta})$:

$$\mathbf{M}(f; \boldsymbol{\lambda}, \boldsymbol{\delta}) = \int_A f(\mathbf{a}) \exp\left[\sum_{i=1}^{n} \lambda_i\varphi_i(\mathbf{a}) + \sum_{j=1}^{m} \delta_j\psi_j(\mathbf{a})\right]P_0(\mathbf{a})d\mathbf{a}. \qquad (5.4)$$

This functional if known helps to compute the values of all regulators. (Notice that this functional is a generalization of the moment generating function (4.2) introduced earlier; namely, $M_0(\boldsymbol{\lambda})=\mathbf{M}(1; \boldsymbol{\lambda}, \mathbf{0})$). Denote

$$\mathbf{S}(t) = \sum_{k=0}^{t} (u_1(S_1(k)), \ldots u_n(S_n(k))),$$

$$\mathbf{H}(t) = \sum_{k=0}^{t} (v_1(H_1(k)), \ldots v_m(H_m(k))).$$

Then the following useful formula is valid:

$$E_0[f\, K_t] = \mathbf{M}(f; \mathbf{S}(t), \mathbf{H}(t)). \tag{5.5}$$

**Theorem 4.** *Let $P_0(\mathbf{a})$ be the initial pdf of the vector-parameter $\mathbf{a}$ for inhomogeneous self-regulated model (5.3), and $\mathbf{M}(f; \boldsymbol{\lambda}, \boldsymbol{\delta})$ be corresponding functional (5.4). Then the total population size and the regulators can be computed recurrently with the help of the system of relations:*

$$N_t = N_0 E_0[K_{t-1}] = N_0 \mathbf{M}(1; \mathbf{S}(t-1), \mathbf{H}(t-1)); \tag{5.6}$$

$$S_i(t) = N_0 E_0[s_i\, K_{t-1}] = N_0\, \mathbf{M}(s_i; \mathbf{S}(t-1), \mathbf{H}(t-1));$$

$$H_j(t) = E_0[h_j\, K_{t-1}]/E_0[K_{t-1}] = \mathbf{M}(h_i; \mathbf{S}(t-1), \mathbf{H}(t-1))/\mathbf{M}(1; \mathbf{S}(t-1), \mathbf{H}(t-1)). \tag{5.7}$$

**Proof.**

Let $N^*_t$, $S^*_i(t)$, $H^*_i(t)$ solve the recurrent system (5.6), (5.7) at given $N_0$, $l(0,\mathbf{a})$, $P_0(\mathbf{a})$, and hence known $\mathbf{S}(0)$ and $\mathbf{H}(0)$. Define

$$w^*_t(\mathbf{a}) = \exp\left[\sum_{i=1}^{n} u_i(S^*_i(t))\varphi_i(\mathbf{a}) + \sum_{j=1}^{m} v_j(H^*_j(t))\psi_j(\mathbf{a})\right],$$

$$K^*_t(\mathbf{a}) = \prod_{k=0}^{t} w^*_k(\mathbf{a}),$$

$$P^*_t(\mathbf{a}) = P_0(\mathbf{a}) K^*_{t-1}(\mathbf{a})/E_0[K^*_{t-1}],$$

$$l^*(t,\mathbf{a}) = P^*_t(\mathbf{a}) N^*_t.$$

Prove that $N^*_t$, $S^*_i(t)$, $H^*_i(t)$, $P^*_t(\mathbf{a})$, $l^*(t,\mathbf{a})$ satisfy system (5.1)-(5.3). This assertion is evidently valid at $t=0$; suppose that it is valid at $t-1$. Then, by supposition, $w^*_{t-1}(\mathbf{a}) = w_{t-1}(\mathbf{a})$ and $K^*_{t-1}(\mathbf{a}) = K_{t-1}(\mathbf{a})$. Next,

$l(t,\mathbf{a}) = l(t-1,\mathbf{a})\, w_{t-1}(\mathbf{a}) = l^*(t-1,\mathbf{a})\, w^*_{t-1}(\mathbf{a}) = P^*_{t-1}(\mathbf{a}) N^*_{t-1}\, w^*_{t-1}(\mathbf{a}) = P_0(\mathbf{a})\, N_0 K^*_{t-1}(\mathbf{a}) = l(0,\mathbf{a})\, K^*_{t-1}(\mathbf{a}) = l^*(t,\mathbf{a});$

$$N_t = \int_A l(t,\mathbf{a})\mathrm{d}\mathbf{a} = \int_A l(0,\mathbf{a})\, K_{t-1}(\mathbf{a})\mathrm{d}\mathbf{a} = \int_A l(0,\mathbf{a})\, K^*_{t-1}(\mathbf{a})\mathrm{d}\mathbf{a} = N_0 E_0[K^*_{t-1}] = N^*_t;$$

$$S(t) \equiv \int_A s(\mathbf{a}) l(t,\mathbf{a})\mathrm{d}\mathbf{a} = \int_A s(\mathbf{a}) l^*(t,\mathbf{a})\mathrm{d}\mathbf{a} = \int_A s(\mathbf{a})\, l(0,\mathbf{a})\, K^*_{t-1}(\mathbf{a})\, \mathrm{d}\mathbf{a} =$$

$N_0 E_0[sK^*_{t-1}] = S^*(t);$

$$H(t) \equiv \int_A h(\mathbf{a}) P(t,\mathbf{a})\mathrm{d}\mathbf{a} = \int_A h(\mathbf{a}) l(t,\mathbf{a})\mathrm{d}\mathbf{a}/N_t = \int_A h(\mathbf{a}) l^*(t,\mathbf{a})\mathrm{d}\mathbf{a}/N^*_t =$$

$$\int_A h(\mathbf{a})\, l(0,\mathbf{a})\, K^*_{t-1}(\mathbf{a})\mathrm{d}\mathbf{a}/N^*_t = E_0[h(\mathbf{a})K^*_{t-1}]/\, E_0[K^*_{t-1}] = H^*(t);$$

$$P_t(\mathbf{a}) = P_0(\mathbf{a})K_{t-1}(\mathbf{a})\,/E_0[K_{t-1}] = P_0(\mathbf{a})K^*_{t-1}(\mathbf{a})\,/E_0[K^*_{t-1}] = P^*_t(\mathbf{a}).$$

Q.E.D.

## 6. The Price equation and the FTNS for selection systems

For the sake of completeness, let us derive the Price equation within the framework of general model (2.1), (2.2). For any sequence $\{s_t, t=0,1,\ldots\}$, denote $\Delta_t s = s_{t+1} - s_t$. Let $z_t(\mathbf{a})$ be a character of an individual with the given vector-parameter $\mathbf{a}$, which can vary with time. Then $E_{t+1}[z_t] = E_t[z_t w_t]/E_t[w]$, $E_{t+1}[\Delta_t z] = E_t[\Delta_t z\, w_t]/E_t[w]$ and

$$E_t[w_t]\,\Delta_t E_t[z_t] = E_t[w_t](E_{t+1}[\Delta_t z] + E_{t+1}[z_t] - E_t[z_t]) =$$

$$E_t[w_t \Delta_t z] + E_t[z_t w_t] - E_t[w_t]E_t[z_t] = \mathrm{Cov}_t[z_t w_t] + E_t[w_t \Delta_t z_t].$$

We obtained the second, or complete, Price equation:

$$\Delta_t E_t[z_t] = (\mathrm{Cov}_t[z_t w_t] + E_t[w_t \Delta_t z])/E_t[w_t].$$

If the character $z$ does not depend on $t$, i.e., $\Delta_t z=0$, then the second Price equation implies the first Price equation, also known as the covariance equation ([29], [21], [23]):

$$\Delta E_t[z] = \mathrm{Cov}_t[w_t, z]/E_t[w_t].$$

If $z=w$, then

$$E_t[w_t]\,\Delta E_t[w_t] = \mathrm{Var}_t[w_t] + E_t[w_t \Delta_t w],$$

which is the FTNS for time-dependent fitness.

If the fitness does not depend on time, i.e. $\Delta_t w=0$, then

$$\Delta E_t[w] = \mathrm{Var}_t[w]/E_t[w],$$

which is the standard form of FTNS.

The Price equations show the connection between the fitness and the selection differential, $\Delta E_t[z_t] = E_{t+1}[z_{t+1}] - E_t[z_t]$, which is an important characteristic of selection. The complete Price equation holds for any particular fitness and any quantitative character; actually, it is a mathematical identity, so it is impossible to "solve" it, i.e., to compute the temporal dynamics of the mean value a particular character without additional information or suppositions.

The theory developed in s.4, s.5 allows us to resolve the problem of dynamical insufficiency of the Price equations and the FTNS. For master model (2.1)-(2.3) and for self-regulated model (5.3), all statistical characteristics of interest could be computed effectively given the initial distribution. In particular, we can compute the mean value of any trait and in this sense to solve the Price equation, the covariance equation, and the equation of the FTNS.

**Proposition 1 (On the complete Price equation)**

i) *For master model* (2.1), (2.3) *with known initial distribution the solution of the Price equation,* $\Delta_t E_t[z_t] = \{Cov_t[z_t w_t] + E_t[w_t \Delta_t z_t]\}/E_t[w_t]$ *is given by the formula*

$$E_t[z_t] = E_0[z_t K_{t-1}]/E_0[K_{t-1}]. \tag{6.1}$$

ii) *For self-regulated model* (5.3) *the solution of the Price equation is given by the formula*

$$E_t[z_t] = \mathbf{M}(z_t; \mathbf{S}(t-1), \mathbf{H}(t-1))/\mathbf{M}(1; \mathbf{S}(t-1), \mathbf{H}(t-1)) \tag{6.2}$$

*The current values of regulators* $\mathbf{S}(t)$, $\mathbf{H}(t)$ *can be computed recursively with the help of Theorem 4, formulas* (5.6)-(5.7).

Indeed, equality (6.1) was proved earlier. Applying formula (5.5), we obtain (6.2).

We supposed in Proposition 1, that all terms in the right hand sides of (6.1) and (6.2) are well defined and finite. Under this supposition, all terms of the Price equation can be computed explicitly:

i) for model (2.1)-(2.3):

$Cov_t[w_t, z_t]/E_t[w_t] = E_0[z_t K_t]/E_0[K_t] - E_0[z_t K_{t-1}]/E_0[K_{t-1}]$,

$E_t[w_t \Delta_t z_t]\}/E_t[w_t] = E_0[(z_{t+1} - z_t) K_t]/E_0[K_t]$;

ii) for model (5.3):

$Cov_t[w_t, z_t]/E_t[w_t] =$

$\mathbf{M}(z_t; \mathbf{S}(t), \mathbf{H}(t))/\mathbf{M}(1; \mathbf{S}(t), \mathbf{H}(t)) - \mathbf{M}(z_t; \mathbf{S}(t-1), \mathbf{H}(t-1))/\mathbf{M}(1; \mathbf{S}(t-1), \mathbf{H}(t-1))$,

$E_t[w_t \Delta_t z_t]\}/E_t[w_t] = \mathbf{M}((z_{t+1} - z_t); \mathbf{S}(t), \mathbf{H}(t))/\mathbf{M}(1; \mathbf{S}(t), \mathbf{H}(t))$.

The FTNS is a particular case of the Price equation under $z_t = w_t$.

**Proposition 2 (On the Fisher FTNS).**

i) *For master model* (2.1)-(2.3) *with known initial distribution the solution of the FTNS equation* $\Delta E_t[w_t] = \{Var_t[w_t] + E_t[w_t \Delta w_t]\}/E_t[w_t]$ *is given by the formula*

$E_t[w_t] = E_0[K_t]/E_0[K_{t-1}]$ .

*If* $M_0(\lambda)$ *is the mgf of the initial distribution of r.v.* $\vartheta = (\varphi_1, \varphi_2, \ldots \varphi_n)$, *then*

$E_t[w_t] = M_0(\mathbf{G}(t))/M_0(\mathbf{G}(t-1))$

where $\mathbf{G}(t) = (G_1(t), \ldots G_n(t))$, $G_i(t) = g_1(s) + \ldots g_t(s)$.

ii) *For self-regulated model* (5.3) *the solution of the FTNS equation is*

$$E_t[w_t] = \mathbf{M}(1; \mathbf{S}(t), \mathbf{H}(t))/\mathbf{M}(1; \mathbf{S}(t-1), \mathbf{H}(t-1)) \tag{6.3}$$

*The current values of regulators $\mathbf{S}(t)$, $\mathbf{H}(t)$ can be computed recursively with the help of Theorem 4.*

All terms of the FTNS equation can also be computed explicitly:

i) for model (2.1)-(2.3):

$Var_t[w_t] / E_t[w_t] = E_0[w_t K_t] / E_0[K_t] - E_0[K_t]/E_0[K_{t-1}]$,

$E_t[w_t \Delta w_t]\}/ E_t[w_t] = E_0[K_{t+1}]/ E_0[K_t] - E_0[w_t K_t]/ E_0[K_t]$,

ii) for model (5.3):

$Var_t[w_t]/E_t[w_t] =$

$\mathbf{M}(w_t; \mathbf{S}(t), \mathbf{H}(t))/ \mathbf{M}(1; \mathbf{S}(t), \mathbf{H}(t)) - \mathbf{M}(1; \mathbf{S}(t), \mathbf{H}(t))/\mathbf{M}(1; \mathbf{S}(t-1), \mathbf{H}(t-1))$,

$E_t[w_t \Delta_t w_t]\}/ E_t[w_t] = \mathbf{M}((w_{t+1} - w_t); \mathbf{S}(t), \mathbf{H}(t))/ \mathbf{M}(1; \mathbf{S}(t), \mathbf{H}(t))$.

Let us stress that formulas (6.2) and (6.3) define the solutions of the Price and FTNS equations for self-regulated selection systems by explicit recurrent procedures; these procedures can be easily realized on computer and sometimes yield solutions in analytical form.

### 7. Applications and Examples.

It is well known that non-linear population models with discrete time (maps) can possess very complex and even counterintuitive behavior depending on the values of model parameters. The main dynamical regimes of the corresponding inhomogeneous models are crucially determined by the behaviors of the original homogeneous models, but have some essentially new interesting peculiarities due to "inner bifurcations", i.e., changing of parameters because of internal dynamics of the system.

### 7.1. Ricker' model

The classical *Ricker'* model $N_{t+1} = N_t \lambda \exp(-bN_t)$ where $\lambda$ and $b$ are positive parameters, takes into account the population size regulation of the reproduction rate. Let us consider the inhomogeneous Ricker' model with a single distributed parameter $a$:

$l(t+1, a) = l(t, a) \lambda_0 a \exp(-bN_t)$,  (7.1)

where the fitness $w_t(a, N_t) = \lambda_0 a \exp(-bN_t)$, $\lambda_0$ is the scaling multiplier and $b>0$ is a constant. According to formulas (3.2)-(3.3), the inhomogeneous Ricker' model takes the form

$N_{t+1} = E_t[w]N_t$,  (7.2)

$E_t[w] = \lambda_0 E_0[a^{t+1}]/E_0[a^t] \exp(-bN_t)$.

Let the initial pdf of *a* be the gamma-distribution with the parameters (*s,k*). Then (as it was noted in s.3, see also [16], Theorem 2.5) $P_t(a)$ is again the $\Gamma$-distribution with the parameters (*s, k+t*), and

$E_t[w] = \lambda_0 (k+t)/s \exp(-bN_t)$,

$N_{t+1} = N_t \lambda_0 (k+t)/s \exp(-bN_t)$.

These formulas show that the coefficient $\lambda_0(k+t)/s$ of the inhomogeneous Ricker' model, which determines the dynamics of the model, increases indefinitely with time. As a result, according to the theory of homogeneous Ricker' model, after a period of monotonic increase cycles of period 2, then of period 4, and then almost all cycles of Feigenbaum's cascade appear and realize as parts of a *single* trajectory, see Fig.1. Hence, after a certain time we observe that the population size begins to oscillate with increasing amplitude. Practically it means that the population goes to extinction with time because there exist points in time when its size happens to be <<1.

The reason of appearance of the remarkable non-classical trajectories shown in Fig.1 is as follows. If the parameter $\lambda_0$ is small and/or *s* is large then the sequence $\{\lambda_0(k+t)/s, t=0,1,...\}$ takes the values close to all bifurcation values of the coefficient of the Ricker' model. It follows that a notable phenomenon, the "almost complete" sequence (with the step $\lambda_0/s$) of all possible bifurcations of the homogeneous Ricker' model is realized within the framework of a unique *inhomogeneous* Ricker' model. The trajectory $\{N_t\}_0^\infty$ in some sense mimics the bifurcation diagram of the plain Ricker' model, see Fig. 1. We would like to emphasize that Figure 1 shows the *trajectory* of the model (7.2) such that to each value of *t* corresponds a single value of $N_t$; for clarification, the enlarged section of the graph is given in Figure 2.

The process of evolution of the population, described by the model, goes through different stages with the speed depending on *s*. Let us explore the evolution of the mean reproduction rate (mean fitness) of the inhomogeneous Ricker' model, $E_t[w]=\lambda_0 E_0[a^{t+1}]/E_0[a^t]\exp(-bN_t)$. Since the fitness depends explicitly on the total population size, it may not increase monotonically with the course of time (in contrast to the plain FTNS and in accordance with the full Price equation). Indeed, this is the case for model (7.1). The dynamics of mean fitness (7.2) together with the total population size for inhomogeneous Ricker' model with the initial Gamma-distribution of the parameter is shown on Fig. 1. We can see that after periods of increase and stable behavior the mean fitness starts to oscillate with increasing amplitude as well as the total population size. A similar phenomenon is observed for any initial distribution of the parameter with unbounded support, e.g., for log-normal distribution.

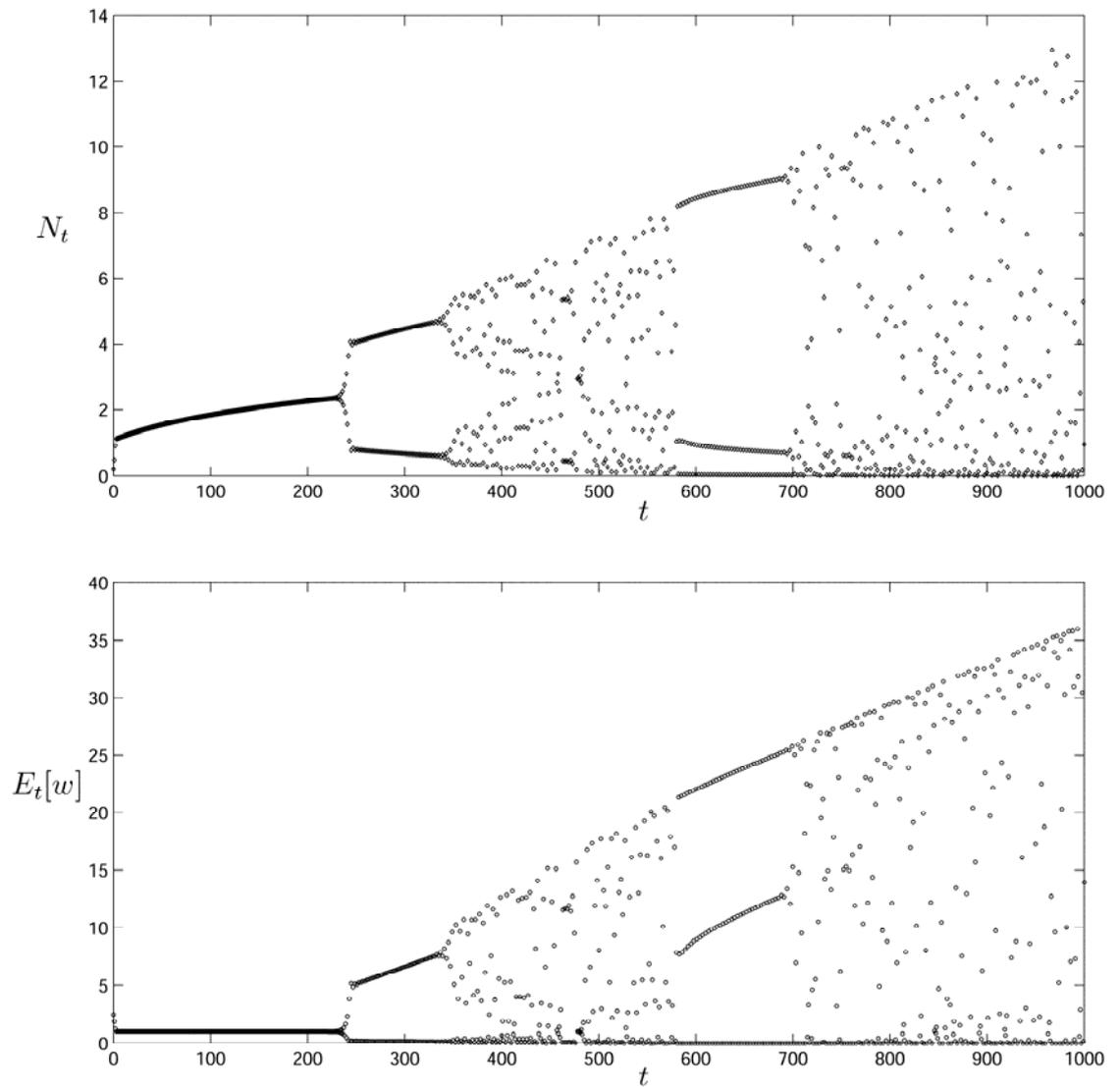

Fig.1. The trajectory of total population size and mean fitness for inhomogeneous Ricker' model with $\Gamma$-distributed parameters $a$ ($\lambda_0$=1, $E_0[a]$ =3, $Var_0[a]$=0.1).

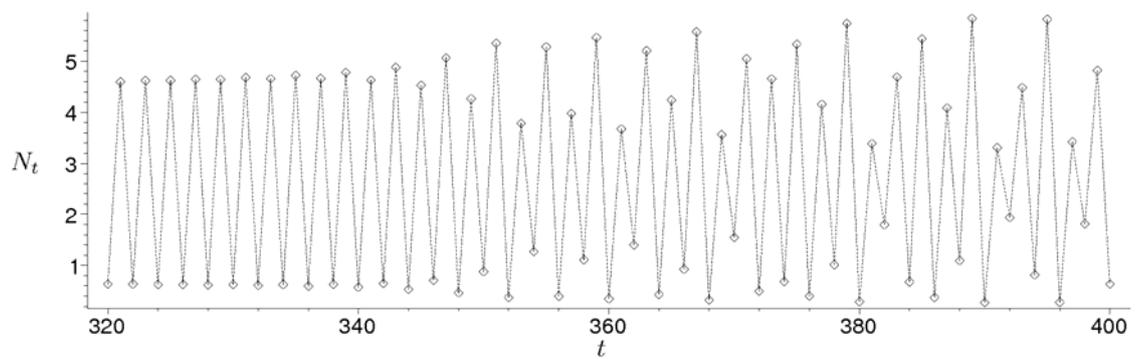

Fig.2. The enlarged section of the model trajectory given in the upper panel of Figure 1.

Another type of behavior is observed if the initial distribution has a bounded support, i.e. the parameter can take any particular value from a *bounded* set A. Let $P_0(a)$ be the Beta-distribution in [0,1] with parameters $\alpha, \beta$. Then $P_t(a)$ is again the Beta-distribution with parameters $\alpha+t, \beta$. Hence,

$E_t[w] = E_t[a]\, g(N) = (t+\alpha)/(t+\alpha+\beta)\, \lambda_0 \exp(-bN_t)$,

$N_{t+1} = N_t\, \lambda_0 E_t[a]\exp(-bN_t) = N_t\, \lambda_0\, (t+\alpha)/(t+\alpha+\beta)\, \exp(-bN_t)$.

Choosing an appropriate value of $\lambda_0$ we can observe any possible behavior of the model as its final dynamics behavior. The following figure 3 illustrates this assertion.

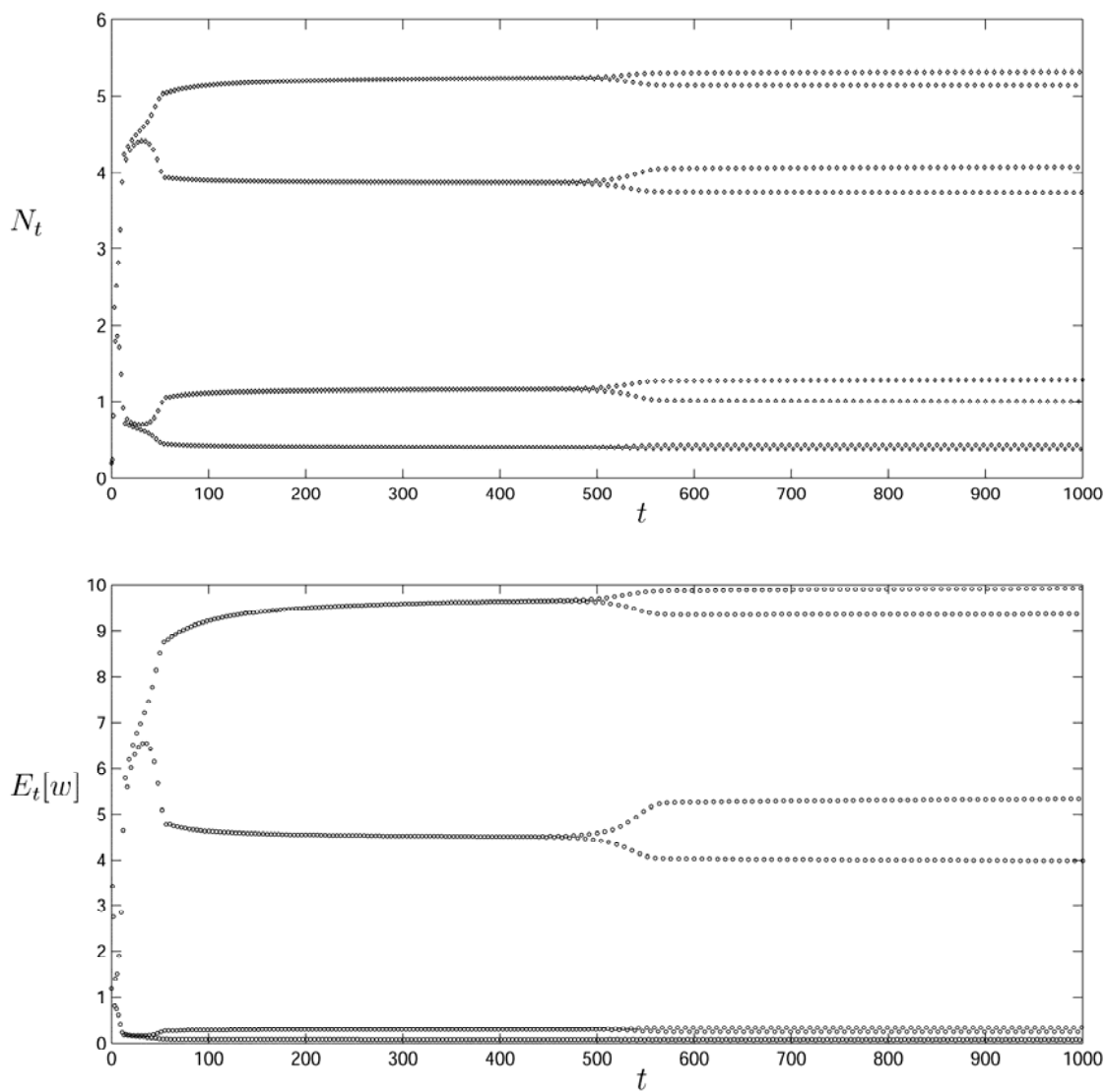

Fig.3. The trajectory of total population size and mean fitness for the inhomogeneous Ricker' model with Beta-distributed parameter ($\lambda_0=14.5$, $E_0[a]=0.1$, $Var_0[a]=0.02$). The final dynamics of the model is 8-cycle.

### 7.2. Logistic model

A well known logistic map is of the form $N_{t+1}=\lambda N_t(1-N_t)$, $0<\lambda<4$ and $0\leq N\leq 1$. Consider the inhomogeneous logistic model

$$l(t+1,a)=\lambda_0 l(t,a)a(1-N_t)$$

where $\lambda_0$=const, $a$ is the distributed parameter; then $w_t(a)=\lambda_0 a(1-N_t)$.

The inhomogeneous logistic model takes the form

$$N_{t+1}=N_t E_t[w], \qquad (7.5)$$

$$E_t[w]=\lambda_0(1-N_t)E_0[a^{t+1}]/E_0[a^t].$$

The model makes sense only if $0<E_0[a^{t+1}]/E_0[a^t]<4$.

Let $P_0(a)$ be the Beta-distribution in [0,1] with parameters $\alpha, \beta$. Then $E_t[w] = \lambda_0(t+\alpha)/(t+\alpha+\beta)(1-N_t)$. Choosing an appropriate value of $\lambda_0 \leq 4$, we can observe (at $t\to\infty$) any possible behavior of the plain logistic model as the final dynamical behaviors of the inhomogeneous logistic model. In particular, at $\lambda_0=4$ almost all cycles of Feigenbaum's cascade appear in the course of time and realize as parts of a *single* trajectory, as a result of the "inner" bifurcations of the inhomogeneous logistic model. Figure 4 illustrates this assertion and also shows a complex behavior of the mean fitness (as distinct from the plain FTNS).

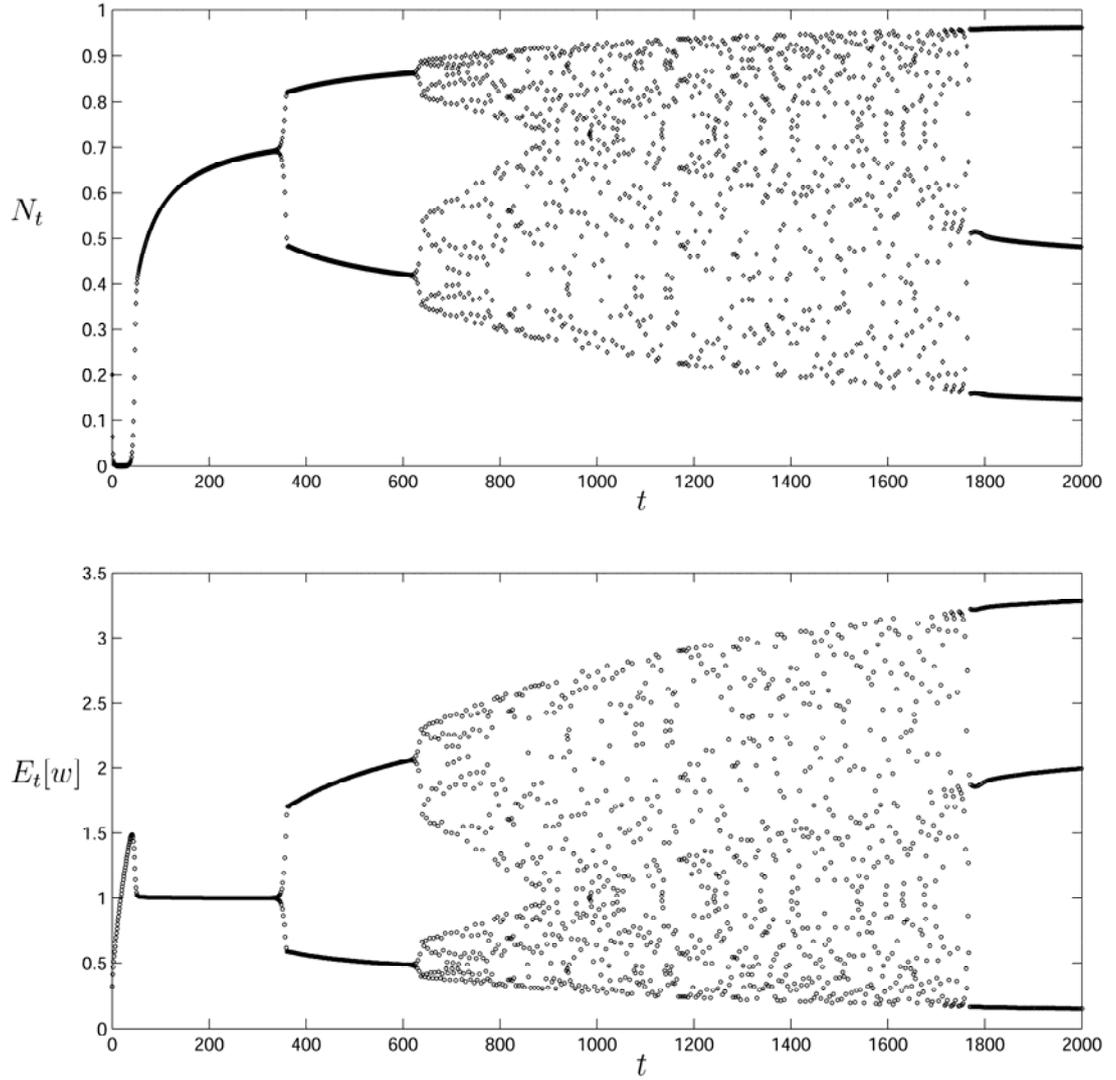

Fig.4. The trajectory of total population size and mean fitness for inhomogeneous logistic model with Beta-distributed parameter $a$ ($\lambda_0=4$, $E_0[a]=0.1$, $Var_0[a]=0.005$).

### 7.3. Ricker' model with two distributed parameters

The Ricker' model takes into account the population size regulation of the reproduction rate by more appropriate way then the logistic map. Consider the inhomogeneous version of the Ricker' model with both distributed parameters $a=\ln\lambda$ and $b$:

$l(t+1;a,b) = l(t; a,b)w(a,b, N_t)$ where $w(a,b, N_t)=\exp[a-bN_t]$.

So $K_t(\mathbf{a})= \prod_{k=0}^{t} w_k(\mathbf{a})=\exp[a(t+1) - b\sum_{k=0}^{t} N_k]$ and

$l(t;a,b) =\exp[a(t+1) - b\sum_{k=0}^{t} N_k] \, l(0;a,b)$.

Comparing with (5.3) we should put for this example $s(\mathbf{a})=1$, $u_1(x)=1$, $u_2(x)=-x$, so that

$S(t)=N_t$, $u_1(S(t))=1$, $u_2(S(t))=-N_t$.

Let $M(\lambda_1, \lambda_2)=\int_A \exp(\lambda_1 a + \lambda_2 b) P_0(a,b) da db$

be the mgf of the initial distribution of parameters $a$ and $b$. Then

$$E_0[K_t] = M(t+1, -\sum_{k=0}^{t} N_k). \qquad (7.3)$$

Applying Theorem 2 we obtain

$$N_t = N_0 E_0[K_{t-1}] = N_0 M(t, -\sum_{k=0}^{t-1} N_k);$$

$$P_t(a,b) = P_0(a,b) \exp[at - b\sum_{k=0}^{t-1} N_k] / M(t, -\sum_{k=0}^{t-1} N_k).$$

These formulas completely solve the inhomogeneous Ricker' model.

The selection differential for the model is $\Delta E_t[w_t] = E_0[K_{t+1}]/E_0[K_t] - E_0[K_t]/E_0[K_{t-1}]$ where $E_0[K_t]$ for given initial distributions can be computed recurrently by formula (7.3).

### 7.4. Selection in Natural Rotifer Community

The mathematical model of zooplankton populations was suggested in (Snell, Serra 1998) and studied systematically in (Berezovskaya et al. 2005). The model depends on the parameters $a$, characterizing the environment quality, and $\gamma$, which is the species-specific parameter. The model takes the form

$$N_{t+1} = N_t w(N_t, a, \gamma),$$

$$w(N_t, a, \gamma) = \exp\{-a + 1/N_t - \gamma/N_t^2\}.$$

The case when only the parameter $a$ is distributed was studied in [17]. Let us consider now the model of a community that consists of different rotifer populations; individuals inside the populations may have different reproduction capacities under constant toxicant exposure. The model is of the form $l(t+1,\mathbf{a})=l(t,\mathbf{a})w_t(\mathbf{a})$, where $\mathbf{a}=(a,\gamma)$ and

$w_t(\mathbf{a})=\exp[-a +1/N_t -\gamma/N_t^2]$,

$$K_t(\mathbf{a})= \prod_{k=0}^{t} w_k(\mathbf{a})=\exp[-(t+1)a +1/\sum_{k=0}^{t} N_k -\gamma/\sum_{k=0}^{t} N_k^2].$$

We have to put in this example (compare to (5.3)) $s(\mathbf{a})=1$, $\varphi_1(\mathbf{a})=a$, $u_1(x)=-1$, $\varphi_2(\mathbf{a})=\gamma$, $u_2(x)=-1/x^2$, so that $S_1(t)=S_2(t)=N_t$, $u_1(S_1(t))=-1$, $u_2(S_2(t))=-1/N_t^2$.

Let $M(\lambda_1,\lambda_2)$ be the mgf of the initial distribution of $a$ and $\gamma$. Then

$$E_0[K_t] = \exp(1/\sum_{k=0}^{t} N_k) \, M(-(t+1), -1/\sum_{k=0}^{t} N_k^2),$$

$$N_t = N_0 \exp(1/\sum_{k=0}^{t-1} N_k) \, M(-t, -1/\sum_{k=0}^{t-1} N_k^2);$$

$$P_t(a, \gamma) = P_0(a, \gamma) \exp[-at - \gamma/\sum_{k=0}^{t-1} N_k^2] / M(-t, -1/\sum_{k=0}^{t-1} N_k^2).$$

These formulas completely solve the inhomogeneous model of rotifer community.

The trajectory $N_t$ has a very complex transition regime from the initial to the final behavior, see Figure 4 (where independent parameters $a, \gamma$ are both Gamma- distributed).

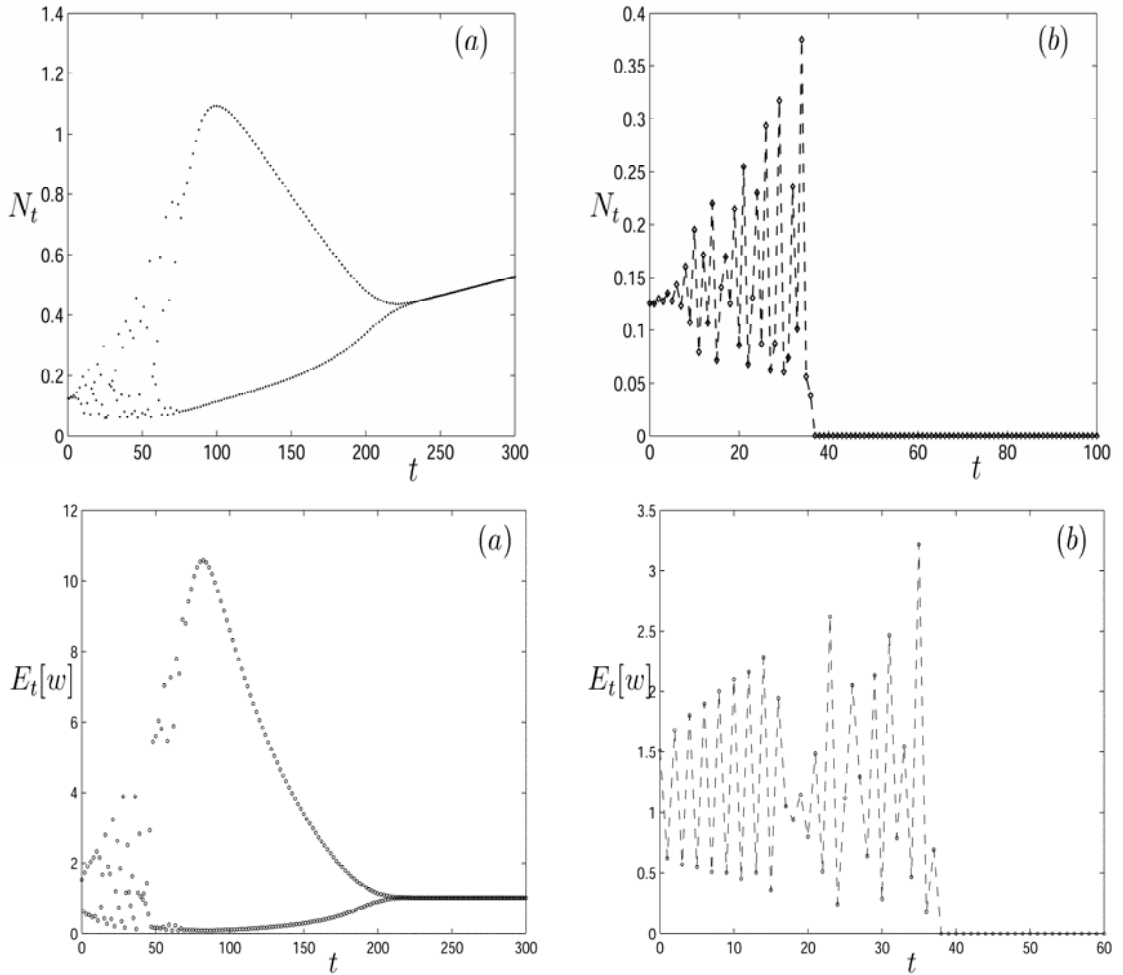

Fig.5. The trajectory of total population size and mean fitness for inhomogeneous rotifer' model with $\Gamma$-distributed parameters $a$; $\gamma=0.044$, $E_0[a]=5.2$. (a) The initial variance is $Var_0[a]=0.035$ and the population reaches the stable asymptotical state. (b) The initial variance is $Var_0[a]=0.026$ and the population goes extinct being trapped in the 0-attracting domain.

**Discussion and Conclusion**

In this paper we tried to make a contribution to the general theory of selection systems with discrete time and develop methods for investigation of the systems evolution in detail. The Fisher Fundamental theorem of natural selection, the Price equations and the Haldane principle are well known general results of the mathematical selection theory. We explore these statements within the framework of a general class of selection systems; we obtain the main results in explicit form as consequences of our investigation of the evolution of the system distribution.

G. Price in the early 70$^{th}$ of the previous century tried to find a general formulation of selection that could be applied to any, not necessary biological, problem of selection and to develop a formal theory. The Price equation was an outstanding contribution to the future theory. It indicates the instant trend of the system dynamics and helps to better understand the connection between the main statistical quantities of the system. This equation is valid in a very general context, but applying it to computing the mean trait seems to create a problem of "dynamical insufficiency". Consequently, the Price equation has little practical value, for it does not allow one to predict changes in the mean of a trait beyond the immediate response. The same is valid for the FTNS, which is a particular case of the Price equation within the framework of selection systems.

The dynamical insufficiency seems to restrict the possibilities of applications of these equations. However, both the Price equation and FTNS are mathematical *identities* and therefore their "solutions" cannot be approached in the same way as the solutions of typical equations. The only way to predict the dynamics of a trait with the help of the Price equation for a long time is to compute all the values in one hand side of the equation *independently* of the other hand side for all time moments of interest. In general, it can be done only if the entire distribution during the total time interval is known or can be computed (and then the Price equation is not necessary).

The examples considered in this paper show the differences in the global dynamics of a selection system, depending on the initial distribution. Note that the Price equation does not depend on the initial distribution, but only on the mean and covariance of the trait and fitness at a given instant. Hence, the Price equation describes a general *instant* property of any selection system, which does not depend on the particular global dynamics of the system. This independence is the reason for the theoretical universality and restricted practical utility of the Price equation taken alone.

The Haldane optimal principle [14] can be considered as one of the first general assertion about selection systems. This principle describes the asymptotical behavior of a population composition; it was generalized in [31], [12] for abstract selection systems (a version of this principle is given here in s.4, (4.3)). Roughly, the composition of the population in *stable* equilibrium is poor: the population is concentrated in the points of global maximum of the mean reproduction coefficient. The system "forgets" all peculiarities of its previous dynamics. The asymptotical composition of general "systems with inherence" driven by selection was explored in mathematical detail in [13]. So, the theory in its recent condition allows one to predict the behavior of selection systems only at the first time step and "at infinity" (if the limit distribution exists and is stable). Remark that the case of indefinite growth of a population similar to the Malthusian model is out of the theory of systems with inherence. Let us emphasize that the current dynamics of the population and its distribution during protracted but finite time intervals is also of interest and, perhaps, is of primary importance in applications.

In this paper we developed the methods that allow us to determine the current distribution of a selection system, which are applicable to a wide class of systems. In particular, effective formulas for computation of the mean of any trait at any time moments are derived at a given initial distribution; these mean values, of course, satisfy to the Price equation and in this sense give its "solution". It is the way to resolve the problem of dynamical insufficiency for the Price equations and for the FTNS. Examples show a complex behavior of the total population size and the mean fitness (in contrast to the plain FTNS) for inhomogeneous populations with the size-dependent fitness.

The developed theory can be applied to a wide class of inhomogeneous population models, which are the maps with distributed parameters. The model behavior may be different and even counter intuitive even for simplest linear maps depending on the initial distribution. Non-linear inhomogeneous models of self-regulated populations show complex dynamical behavior; their trajectories may mimic parts of the model bifurcation diagram. So the models can have extremely complex transitional regime from the initial to the final behavior. For instance, all cycles of Feigenbaum's cascade and even chaotic behavior may appear in the course of time and realize as parts of the *single* trajectory due to the "inner" bifurcations of the inhomogeneous model. Additionally, any possible behavior of corresponding "homogeneous" model may be observed as a final behavior (at $t \to \infty$) with appropriate initial distribution.

We hope that this paper may be useful for understanding of the dynamic peculiarities of inhomogeneous maps and the crucial role of the initial distributions; we hope that theorems and methods presented here can help to explore (analytically and numerically) inhomogeneous self-regulated population models with discrete time, which appear in different areas of mathematical biology.

**Acknowledgment**. The author thanks Dr. A. Kondrashov and Dr. E. Koonin for valuable discussions, Dr. A. Novozhilov for preparation the figures, and I. Kareva for help in preparation of the manuscript.